\documentclass[conference,a4paper]{IEEEtran}
\usepackage[utf8]{inputenc}
\usepackage[T1]{fontenc}
\usepackage{cite}
\usepackage{amsmath,amssymb,amsfonts}
\usepackage{algorithm}
\usepackage{algorithmic}
\usepackage{graphicx}
\usepackage{textcomp}
\usepackage{xcolor}
\usepackage[hyphens,spaces]{url}
\usepackage{hyperref}
\usepackage{booktabs}
\usepackage{tabularx}
\usepackage{tikz}
\usetikzlibrary{positioning,arrows.meta,calc,fit,backgrounds}
\usepackage{listings}

\def\BibTeX{{\rm B\kern-.05em{\sc i\kern-.025em b}\kern-.08em
    T\kern-.1667em\lower.7ex\hbox{E}\kern-.125emX}}

\newcommand{\circled}[1]{\textsuperscript{\textcircled{\scriptsize #1}}}

\title{Bonded Recourse for Smart-Contract Settlement of Compensable Agent Side Effects}

\author{%
    \IEEEauthorblockN{Laurent Bindschaedler\IEEEauthorrefmark{1}, Quentin Botha\IEEEauthorrefmark{2}, Christoph Siebenbrunner\IEEEauthorrefmark{2}}
    \IEEEauthorblockA{\IEEEauthorrefmark{1}Max Planck Institute for Software Systems\\
    bindsch@mpi-sws.org}
    \IEEEauthorblockA{\IEEEauthorrefmark{2}Research Institute for Cryptoeconomics, Vienna University of Economics and Business\\
    \{quentin.botha, christoph.siebenbrunner\}@wu.ac.at}
}

\def\tablerqone{%
  postgres & atomix & 6796 & 1796 & 0 & 5000 \\
  postgres & auth\_plus\_local & 6796 & 1796 & 0 & 5000 \\
  postgres & centralized & 5464 & 1796 & 1198 & 2470 \\
  postgres & human & 3967 & 0 & 0 & 3967 \\
  postgres & none & 6796 & 0 & 0 & 6796 \\
  postgres & oap & 6796 & 0 & 0 & 6796 \\
  postgres & recourse & 5464 & 1796 & 1198 & 2470 \\%
}

\def\tablerqfour{%
  atomix & 6057 & --- & 0.00 & 0.00 \\
  centralized & 3511 & 1.00 & 0.00 & 0.67 \\
  oap & 8220 & --- & 0.00 & 0.00 \\
  recourse & 3511 & 1.00 & 0.00 & 0.67 \\%
}

\def\tablerqfive{%
  BND\_DOWNSTREAM\_LEAK & 0.00 & 0.00 & 0.00 & 1.00 \\
  BND\_OUT\_OF\_CAP\_LOSS & 0.00 & 0.00 & 0.00 & 1.00 \\
  CMP\_SUBJECTIVE\_RULE & 0.00 & 0.00 & 1.00 & 0.00 \\
  NONE & 0.00 & 1.00 & 0.00 & 0.00 \\
  OBS\_AMBIGUOUS\_RESOURCE & 0.00 & 0.00 & 1.00 & 0.00 \\
  OBS\_HIDE\_PRESTATE & 0.00 & 0.00 & 1.00 & 0.00 \\
  OBS\_REMOVE\_PROVIDER\_SIG & 0.00 & 0.00 & 1.00 & 0.00 \\
  REC\_INVALID\_ARTIFACT & 0.00 & 0.00 & 1.00 & 0.00 \\
  REC\_PARTIAL\_FAILURE & 0.00 & 0.00 & 1.00 & 0.00 \\
  REC\_STALE\_SNAPSHOT & 0.00 & 0.00 & 1.00 & 0.00 \\%
}

\def\tablerqsix{%
  register                      &     578{,}185 &     578{,}185 \\
  postBond                      &     142{,}314 &     142{,}314 \\
  commitReceipts                &     208{,}554 &     208{,}542 \\
  submitClaim                   &     410{,}603 &     410{,}591 \\
  assertResidualLoss            &     412{,}211 &     412{,}187 \\
  dispute                       &           --- &      66{,}221 \\
  settle (finalize + payout)    &     102{,}230 &           --- \\
  escalate                      &           --- &     142{,}143 \\
  giveRuling (rule + payout)    &           --- &      99{,}668 \\
  total listed call-gas         & 1{,}854{,}097 & 2{,}059{,}851 \\%
}

\providecommand{\tablekcatalogue}{%
  k-postgres-update & postgres-table & \$1{,}000 & \$10{,}000 & 552{,}446 \\
  k-postgres-ddl & postgres-table & \$5{,}000 & \$50{,}000 & 552{,}450 \\
  k-git-pushforce & git-ref & \$500 & \$5{,}000 & 552{,}532 \\
  k-cloud-ssm & cloud-resource & \$1{,}500 & \$15{,}000 & 552{,}610 \\
  k-cloud-rds-snapshot & cloud-resource & \$750 & \$7{,}500 & 552{,}646 \\
}

\def\tablerqfiveadv{%
  NONE & 0.00 & 0.00 & 3782 & 3782 \\
  DISHONEST\_CLAIMANT & 0.00 & 1.00 & 5272 & 0 \\
  ABSENT\_CHALLENGER & 0.00 & 0.00 & 18853 & 18853 \\
  BOND\_STRESS & 0.00 & 0.00 & 4036 & 4036 \\
  COLLUDING\_CHALLENGERS & 0.00 & 0.58 & 4676 & 3782 \\%
}

\begin{document}
\pagestyle{empty}
\maketitle

\begin{abstract}
Autonomous agent runtimes execute tool actions that mutate databases, repositories, and cloud services across organizational boundaries. Authorization and local compensation cover pre-action admission and in-runtime rollback, but neither settles the residual harm left after a permitted action fails. We design \emph{Recourse}, a smart-contract settlement protocol for compensable agent side effects that binds each admitted action to scope, recovery, evidence, payout, and collateral. Recourse separates \emph{ex ante} eligibility from \emph{ex post} objective settleability: typed receipts make objective residual claims computable under an optimistic-oracle challenge pattern, while subjective or incomplete claims route to ERC-792 arbitration or exclusion. We implement the contract suite, deploy it on Base Sepolia, build adapters against Postgres, Git, and cloud-compatible local sandboxes, and evaluate the system on a deterministic harness, sandbox traces, adversarial sweeps, and property-based fuzzing. Against authorization-only and local-compensation baselines, bonded coverage cuts uncompensated harm. The on-chain tier supplies neutral custody, public challenge, non-cooperative payout, and portable history under cross-organizational trust assumptions.
\end{abstract}

\begin{IEEEkeywords}
autonomous agents, blockchain, smart contracts, recourse, arbitration, optimistic oracles, agent safety
\end{IEEEkeywords}

\section{Introduction}
\label{sec:intro}

Autonomous agent runtimes now sit on production control paths. They update databases, rewrite repositories, modify cloud configuration, and call administrative APIs across organizational boundaries. Benchmarks such as SWE-bench~\cite{Jimenez2024SWEBench} and the agents built for them~\cite{Yang2024SWEAgent} already exercise these tool calls in autonomous workflows.

A recent Replit agent deleted a production database during a code freeze~\cite{Lemkin2025ReplitIncident}. Pre-action permissioning decides which calls may run. Operators still need a mechanism that permits bounded-risk actions and settles the residual harm an admitted action leaves across separate control planes.

Existing mechanisms cover adjacent decisions. Open Agent Passport (OAP)~\cite{Uchibeke2026BeforeToolCall} applies deterministic authorization before tool execution. Transactional runtimes such as Atomix~\cite{Mohammadi2026Atomix} compensate leaked effects within a runtime workflow boundary. Task-payment systems such as TessPay~\cite{Goenka2026TessPay} settle payment for completed work. Human approval suppresses autonomy, and centralized escrow concentrates custody and release in one platform. None supplies the record needed to invoke recovery and compute payout after an admitted action fails.

The unresolved case appears after an allowed action executes, triggers recovery, and still leaves bounded residual harm. Examples include over-broad deletes with snapshot restore, force-pushes with manual repair, and autoscaling changes with capped excess spend. These failures stay bounded when the contract fixes scope, recovery, evidence, and payout before execution, even when full reversibility is unavailable.

We design Recourse for this compensable class of agent side effects. Recourse uses a two-stage admission/settlement boundary. Before execution, an action is \emph{recourse-eligible} when a machine-readable contract fixes the resource scope, recovery operator, covered residual-loss class, payout rule, and required bond. After execution, an instance is \emph{objectively settleable} when typed effect and recovery receipts compute the payout from contract fields alone. When receipts cover effect and recovery, Recourse computes payout as a bounded function of contract fields. Incomplete, subjective, or excluded cases route to arbitration or exclusion.

Recourse keeps agent execution off-chain. The chain supplies the cross-organizational settlement substrate: vault rules govern collateral release, third-party watchers can dispute from on-chain commitments, the vault pays out even if the operator refuses to act, and claim outcomes outlive any individual operator~\cite{Goenka2026TessPay,Kleros2026Escrow,Kleros2026Oracle}: neutral custody, public challenge, non-cooperative payout, and portable history. Recourse covers harms parties can bound and evidence before execution. Data exfiltration, public leakage, irreversible transfers, and reputation damage stay outside it.

Recourse covers one decision: bounded residual loss for an admitted action under a concrete recourse contract, leaving pre-action authorization, workflow atomicity, and agent-quality guarantees to other mechanisms.

Recourse and a centralized-escrow control share the per-claim payout rule and settle identically in covered loss. Both cut uncompensated harm by 49.8\,\%\,$\pm$\,12.6\,pp (paired, 95\% CI across 10 seeds) versus an authorization-and-local-recovery baseline on a deterministic Postgres harness. The two systems differ only on trust-model guarantees.

We make three contributions.
\begin{itemize}
  \item The eligibility/objective-settleability boundary separating compensable agent side effects from arbitrated and excluded harms, pinned to seven machine-checkable predicates over scope, evidence, recovery, residual, exclusion, computability, and collateral: a boundary and routing discipline, not a new cryptographic primitive (Section~\ref{sec:settleability}).
  \item An instantiation of that boundary composing existing settlement primitives (UMA OOv3 for optimistic assertion, ERC-792/1497 for arbitration) around a recovery-first bond vault and a typed-receipt evidence schema, so routing across optimistic, arbitrated, and excluded paths reduces to a single predicate evaluation (Sections~\ref{sec:system}--\ref{sec:settlement}).
  \item A prototype and evaluation: a bytecode-verified Base Sepolia deployment, end-to-end on-chain settlement, adapters for three provider-compatible sandboxes, an adversarial sweep over four threat classes, and property-based fuzzing (Sections~\ref{sec:implementation}--\ref{sec:evaluation}).
\end{itemize}

\section{Background and Motivation}
\label{sec:taxonomy}

Autonomous tool actions differ in how their effects can be repaired and measured. A post-failure mechanism must distinguish three cases: (i)~actions the runtime can undo locally, (ii)~actions that leave bounded residual loss, and (iii)~actions whose harm escapes any pre-execution bound.

\subsection{Effect Classes and Post-Failure Gap}

We distinguish three classes of autonomous side effects. \emph{Reversible} actions have reliable runtime or provider undo paths (e.g., versioned file writes, database updates with trustworthy undo logs). These need replay discipline alone. \emph{Compensable} actions cause scope-closed harm that agreed evidence can measure as recovery cost or capped residual loss, and that posted collateral can reimburse (e.g., over-broad delete with snapshot restore, repository rewrite with manual repair, misconfiguration with bounded cloud spend). \emph{Excluded} actions (exfiltration, public leaks, irreversible transfers, physical-world effects) have a downside outside both credible recovery and credible advance measurement. They need hard approvals.

Residual harm after an admitted compensable action fails sits outside all three. \emph{Effect class} is an admission test, while \emph{settlement route} is the post-failure choice: \emph{reversible} actions (local compensation exists) stay local via rollback or abort, \emph{compensable} actions (contract, bond, bounded loss) get recovery plus settlement, optimistic if objective and arbitrated if covered but non-objective, and \emph{excluded} actions (unbounded or non-recoverable) hit a hard gate or approval, outside the model. Arbitration is a failure path for compensable actions, separate from the three admission classes.

\subsection{Threat Model}

We consider three partially mistrusting roles: a customer affected by agent actions, an operator running the agent, and a tool or infrastructure provider that can expose state evidence. The operator may be honest but careless, under-bonded, or economically motivated to delay compensation, and a claimant may exaggerate residual loss. Watchers can be unavailable or economically inactive, and tool providers may expose incomplete or weak receipts.

We assume smart-contract integrity on the underlying chain. Chain-level consensus failures stay outside our scope.

We grade receipt provenance by who can attest a receipt or replay it. Objective settlement requires provider-signed receipts, snapshot references, or independently checkable restore artifacts. Weak evidence routes to review. Five failure cases drive routing: (i)~weak contracts fail admission, (ii)~missing or ambiguous evidence breaks observability, (iii)~stale snapshots or failed compensators break executed recovery, (iv)~uncapped downstream effects break boundedness, and (v)~payout disputes break computability. Public disclosure, exfiltration, irreversible transfer, and uncapped downstream effects stay outside the model regardless of any payout cap written into the contract. A provider-signed or independently replayable receipt supports objective settlement. Every other receipt is operator-attested evidence, and it routes the residual claim away from objective settlement.

\section{Eligibility and Settleability}
\label{sec:settleability}

Recourse makes the two decisions from Section~\ref{sec:taxonomy} explicit: whether the pre-execution (ex ante) contract covered the action, and whether post-recovery receipts make the failed instance \emph{settleable}.

For one recourse contract, $\kappa = (a, x, \rho, L, P, B, S, \Delta)$ where $a$ is the action template, $x$ the resource scope, $\rho$ the registered recovery operator, $L$ the covered residual-loss class, $P$ the payout rule, $B$ the locked bond, $S$ the evidence schema, and $\Delta$ the deadlines (recovery, submission, challenge). The runtime admits $\kappa$ only if it verifies, before execution, that (i)~scope $x$ is closed under the action's side-effect channels; (ii)~$\rho$ is available under freshness/retention/authorization preconditions; (iii)~$S$ names the evidence sources at the required provenance; (iv)~$L$ admits recoverable measurable harm; (v)~$P$ is total over schema-valid recovery outcomes; and (vi)~the bond covers the contract's maximum aggregate exposure.

For one failed instance under $\kappa$, $\sigma = (\kappa, o, r)$ where $o$ is the effect evidence (pre/post-state digests, provider receipts, timestamps) and $r$ is the recovery evidence (restore outcome, rollback reference, schema-defined residual measurements). $\kappa$ records the pre-execution promise. $\sigma$ records the executed action, recovery, and residual outcome.

$\sigma$ is \emph{objectively settleable} when four conditions hold jointly: (i)~$o$ establishes the action and resource from typed receipts and conforms to scope $x$; (ii)~$\rho$ is invokable and $r$ records the attempted outcome; (iii)~the receipts and schema restore, measure, cap, or rule out each consequence in the contracted loss class $L$; and (iv)~$P(o,r)$ maps the evidence to a bounded payout from contract fields alone. The contract excludes consequences outside $L$, including exfiltration, irreversible transfers, and uncapped downstream effects. Concretely:

{\footnotesize
\begin{align*}
\textsc{Settleable}(\sigma) ={}&
\textsc{Eligible}(\kappa)\\
& \wedge \textsc{WithinScope}(o,x)\wedge
\textsc{EvidenceOK}(o,r,S)\\
& \wedge \textsc{RecoveryOK}(r,\rho,\Delta)\\
& \wedge \textsc{ResidualOK}(o,r,L)\\
& \wedge \textsc{NoExcluded}(o,r)\\
& \wedge \textsc{Computable}(P,o,r)\\
& \wedge \textsc{CollateralOK}(P,o,r,B).
\end{align*}
}

Eligibility is a pre-execution admission test on $\kappa$, while objective settleability is a post-execution routing test on $\sigma$. When every settleability predicate holds, the failed action stays optimistic. Failures of evidence, recovery, computability, or collateral predicates route covered below-cap claims to arbitration. Failures of scope, residual, or exclusion predicates move the instance outside the objective Recourse path.

Table~\ref{tab:predicates} maps each predicate to contract fields, receipt checks, and settlement route. Algorithm~\ref{alg:route} gives the router.

\begin{algorithm}[t]
\caption{Route for a failed instance $\sigma=(\kappa,o,r)$ under a recourse contract.}
\label{alg:route}
\scriptsize
\begin{algorithmic}[1]
\IF {$\neg\textsc{Eligible}(\kappa)$} \RETURN \emph{ineligible} \ENDIF
\IF {$\neg(\textsc{WithinScope} \wedge \textsc{ResidualOK} \wedge \textsc{NoExcluded})$} \RETURN \emph{excluded} \ENDIF
\IF {$\textsc{EvidenceOK} \wedge \textsc{RecoveryOK} \wedge \textsc{Computable} \wedge \textsc{CollateralOK}$} \RETURN \emph{optimistic} \ENDIF
\RETURN \emph{arbitrated} \COMMENT{contract default within $\kappa.P$, $\kappa.L$}
\end{algorithmic}
\end{algorithm}

\begin{table}[t]
\caption{Settleability predicates and the corresponding Postgres over-delete checks. The example action is a \texttt{DELETE} within a named schema with a snapshot-restore compensator.}
\label{tab:predicates}
\centering
\scriptsize
\begin{tabularx}{\columnwidth}{@{}p{1.15cm}p{1.05cm}p{1.5cm}p{0.9cm}X@{}}
\toprule
\textbf{Predicate} & \textbf{Verified by} & \textbf{Receipt evidence} & \textbf{Settlement route} & \textbf{Postgres check} \\
\midrule
WithinScope & runtime & $o$ resource id, query hash & excluded & \texttt{schema} prefix matches $x$ \\
EvidenceOK & runtime + challenger & $o,r$ signatures, digests & arbitrated & L1 provider receipt + signed schema \\
RecoveryOK & runtime + challenger & $r$ outcome, timestamps & arbitrated & restore artifact by $\Delta_\rho$ \\
ResidualOK & contract & $r$ residual measurement & excluded & rows-lost within $L$, exfiltration absent \\
Computable & contract & $P(o,r)$ inputs typed & arbitrated & rowcount $+$ downtime, schema-valid \\
CollateralOK & contract & vault state & arbitrated & locked bond at least $P_{\max}$ \\
\bottomrule
\end{tabularx}
\end{table}

\section{System Design}
\label{sec:system}

\begin{figure*}[!t]
\centering
\includegraphics[width=0.78\textwidth]{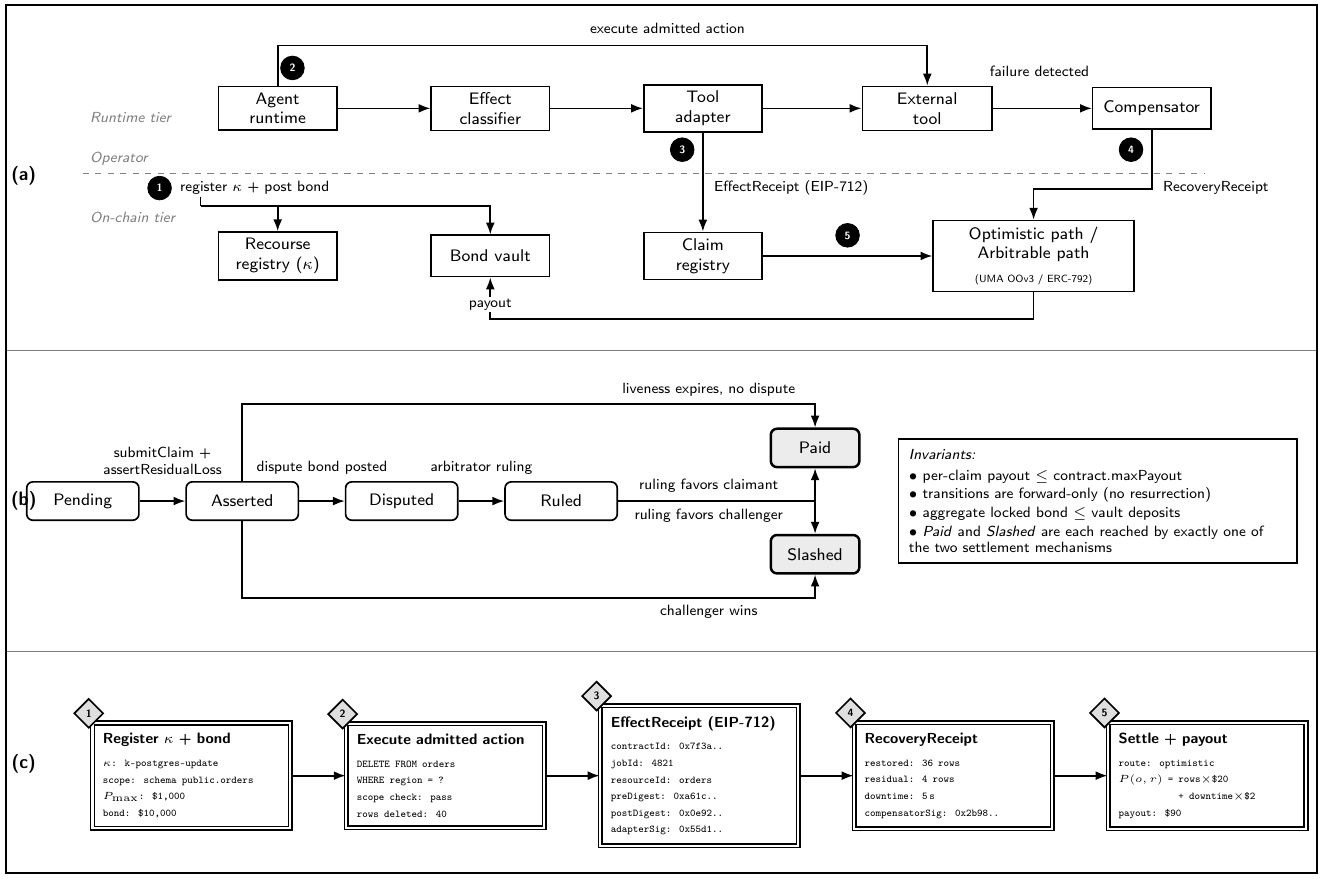}
\caption{Recourse protocol and Postgres worked example. (a) Runtime and on-chain tiers for phases \circled{1}--\circled{5}. (b) Forward-only claim state machine. (c) Postgres over-delete $\kappa$ with concrete receipt and payout fields.}
\label{fig:architecture}
\end{figure*}

Recourse splits responsibilities across a runtime tier (agent, effect classifier, tool adapter, compensator) and an on-chain tier (recourse contract, bond vault, claim registry, optimistic-oracle path, and arbitrator path). Fig.~\ref{fig:architecture}(a) shows the two tiers and the five-phase protocol. The operator registers $\kappa$ and posts bond (\circled{1}). The agent executes an admitted action (\circled{2}). The adapter commits an \emph{EffectReceipt}, a signed record of the tool action (\circled{3}). The compensator commits a \emph{RecoveryReceipt}, a signed record of restore outcome and residual measurement, to the claim registry (\circled{4}). The claim then settles via the optimistic path or escalates to arbitration, with payout released from the bond vault (\circled{5}). Panel (b) is the claim state machine, and panel (c) walks the Postgres over-delete scenario through the same phases with concrete numbers.

\subsection{Recourse Contracts and Bond Vault}
\label{sec:action-contracts}

A recourse contract operationalizes $\kappa$ from Section~\ref{sec:settleability}. Eligibility must be machine-checkable before execution, and payout must compute from contract fields alone. The operator registers one contract per \emph{(action template, resource scope)} pair. The schema pins action and scope, effect class, payout schedule and cap, minimum bond and collateral asset, compensator id, evidence schema with provenance level and retention, recovery freshness preconditions, and claim/liveness/dispute deadlines. Contract quality determines admission. When $L$ is too broad, $P$ too vague, or $S$ unable to produce objective evidence at the required provenance, eligibility fails before execution.

The operator (or a third-party surety) posts collateral into a \emph{BondVault}, the smart contract that locks and releases recourse collateral, separate from any service-fee escrow. The vault removes unilateral operator or customer control over release: operator-controlled release reverts to support tickets, while customer-controlled release can block legitimate release. This is the credible-neutrality role Recourse needs for action-level settlement~\cite{Buterin2020CredibleNeutrality}. The BondVault's neutrality covers the locked collateral. Off-chain evidence truthfulness, provider cooperation, and losses above the posted cap stay outside its guarantee. Admission tracks aggregate exposure: the locked bond must cover open exposure $E_{\mathrm{open}}$ plus the contract's maximum payout $P_{\max}$, the collateral-adequacy condition from Section~\ref{sec:settleability}.

\subsection{Receipts and Recovery-First Workflow}

On execution the adapter emits a signed EffectReceipt binding contract id, job id, intent/parameter hashes, resource id, pre/post-state digests, timestamp, and adapter signature (plus \texttt{providerSig} at L1). Fig.~\ref{fig:architecture}(c) card~\circled{3} shows these fields in the worked example. Recourse groups receipt provenance, the source and checkability of each receipt, into four levels (Table~\ref{tab:provenance}) and admits $\textsc{EvidenceOK}$ at the level the contract declares. Recourse assumes no TEE anywhere in the stack: trust in a receipt rests on key custody at whichever provenance level the contract declares. An L2 receipt is only as trustworthy as the adapter's signing key, so it routes to arbitration rather than optimistic settlement (Table~\ref{tab:provenance}). An L1 receipt requires the provider's own \texttt{providerSig} instead, which moves trust from the operator-controlled adapter to the infrastructure provider. An L0 receipt removes key trust entirely: any node can replay the action against public state.

\begin{table}[t]
\caption{Receipt provenance levels and settlement routes. \textsc{EvidenceOK} checks the receipt against the level declared in $\kappa$.}
\label{tab:provenance}
\centering
\scriptsize
\begin{tabularx}{\columnwidth}{@{}p{0.45cm}p{1.5cm}p{1.5cm}p{1.55cm}X@{}}
\toprule
\textbf{Lvl} & \textbf{Signer} & \textbf{Verifier} & \textbf{Availability} & \textbf{Allowed route} \\
\midrule
L0 & public replayer & any node from public state & on-chain digest & optimistic \\
L1 & provider & any holder of provider key & retention at least $\Delta$ & optimistic \\
L2 & adapter / compensator & operator-known key & evidence store & arbitrated \\
L3 & claimant narrative & dispute review & best-effort & contract default \\
\bottomrule
\end{tabularx}
\end{table}

A detected failure invokes the registered compensator before any payout path executes. The compensator emits a RecoveryReceipt containing the compensator id, restore artifact, outcome (success, partial, failure), schema-defined residual measurements with evidence source, and recovery timestamps. Together with the effect receipt, it forms $r$ and supplies the inputs to $P(o,r)$. Recovery runs before payout because a compensator that restores first leaves payout to cover the residual.

A slash-first protocol collapses action recourse into generic staking: a free-form residual-damage estimate lacks the fields an objective recovery receipt needs at any provenance level, so it forces the instance out of the optimistic path. A missing schema-valid receipt by $\Delta_\rho$ freezes the bond and triggers the contract's default or arbitration route. The compensator must be scope-closed: its side effects must fall under the covered loss class $L$ and payout rule $P$, or eligibility fails.

\section{Claim Settlement}
\label{sec:settlement}

Once the claimant submits commitments to effect and recovery receipts, the bond vault releases funds through a settled claim. Algorithm~\ref{alg:route} selects the route. An objective claim posts an assertion bond and stays open for a liveness window, a challenge period during which any counterparty may dispute by posting a dispute bond. This follows the optimistic-bond-with-challenge pattern of UMA OOv3, an optimistic-oracle protocol for challengeable assertions~\cite{UMA2026Oracle,UMA2026CustomBondLiveness}. The challenge window does not block the agent: the runtime tier executes and admits actions off-chain before any claim exists, and $\Delta$'s liveness and dispute deadlines govern only the settlement instance $\sigma$, not the agent's next action. The vault admits a new action under the same $\kappa$ if the bond covers open exposure plus the new action's maximum payout (Section~\ref{sec:action-contracts}). An operator who keeps the vault collateralized keeps running while earlier claims sit in their challenge window. The agent's exposure is bounded by the eventual payout, independent of the oracle's resolution latency.

Objective settleability requires asserted facts to remain checkable against signed receipts and contract fields. An uncontested optimistic settlement relies on the challenge assumption. Independent proof still comes from the receipts themselves. Failed predicates take distinct routes (Table~\ref{tab:predicates}). Missing evidence routes to arbitration when $\kappa$ defines an evidentiary default. Computability failures route to arbitration within cap and covered loss class. Boundedness failures and excluded harm leave the objective Recourse path. Recourse uses the ERC-792 arbitrable/arbitrator split with ERC-1497 evidence submission, the standard interface~\cite{Lesaege2017ERC792,Vitello2018ERC1497}.

\textbf{Griefing.} A malicious challenger cannot drain the operator's bond simply by escalating to arbitration. ERC-792 arbitration fees are paid by the party requesting arbitration, and Kleros's fee-shifting rules reimburse the winning side from the losing side's deposit: a challenger who disputes a valid claim and loses pays their own arbitration fee, not the operator's. Frivolous disputes cost the challenger. The claimant's bond stays intact. We inherit this guarantee from the ERC-792/Kleros deployment. We do not verify it independently~\cite{Lesaege2017ERC792,Kleros2026Escrow,Kleros2026Oracle}.

A trusted platform can hold escrow and execute payouts with lower complexity. We use the on-chain tier for the cross-organizational trust model where custody, challenge, payout, and history must stay distributed~\cite{Goenka2026TessPay,Kleros2026Escrow,Kleros2026Oracle,Lesaege2017ERC792,Vitello2018ERC1497}. The chain provides \emph{neutral custody}, \emph{public challenge}, \emph{non-cooperative payout}, and \emph{portable history}: centralized escrow fails on all four, since an operator who refuses payout after settlement simply keeps the funds, the platform can censor dispute submission, evidence need not outlive the operator, and custody sits with the operator rather than a neutral vault when operator and customer disagree. Recourse passes all four by construction.

\textbf{Watcher incentive.} The optimistic path requires a rational challenger. For assertion bond $b_a$, dispute bond $b_d=\gamma b_a$, and inspection cost $c$, a watcher's expected value from challenging is $p \cdot b_a - (1-p) \cdot \gamma b_a - c$. The rational-challenge threshold is $p^* = (\gamma + c/b_a)/(1+\gamma)$. With matched bonds ($\gamma=1$) and $c \ll b_a$, $p^* \to 0.5$. Lowering $\gamma$ reduces the threshold but opens griefing, so $\kappa$ parameterizes both bonds and lets operators tune $\gamma$ against the expected distribution of false claims. The \textsc{ABSENT\_CHALLENGER} loss (RQ3) is the case where no party has $p > p^*$ given $c$.

This is the same rational-watcher assumption underlying Truebit and optimistic-rollup fraud proofs~\cite{Teutsch2023ScalableVerification,Kalodner2018Arbitrum,OPLabsFaultProof}: security holds only while at least one economically motivated party monitors each class of claim. Recourse exposes $\gamma$ and $b_a$ as the only levers and does not provision or subsidize watchers. Operators should widen the bond spread for high-value or thinly monitored scopes to keep $p^*$ low, but low-value or niche scopes stay exposed to \textsc{ABSENT\_CHALLENGER} until a paid watcher market exists.

\section{Prototype Implementation}
\label{sec:implementation}

The on-chain tier is a Solidity contract suite (registry, bond vault, claim registry, optimistic path, arbitrable path) with USDC-denominated bonds, deployed to a public testnet. Objective claims dispatch to UMA OOv3~\cite{UMA2026Oracle,UMA2026CustomBondLiveness}, while non-objective claims dispatch to an ERC-792/1497 arbitrable interface with Kleros~\cite{Lesaege2017ERC792,Vitello2018ERC1497,Kleros2026Escrow,Kleros2026Oracle}. The runtime tier uses per-provider adapters for Postgres, Git, and cloud-compatible local sandboxes, paired with a classifier that loads $\kappa$, evaluates eligibility, dispatches admitted actions, and invokes the registered compensator. Receipts serialize as EIP-712 typed-data signatures. Large artifacts sit in a content-addressed evidence store, while content hash, schema id, retention, and policy hit calldata stay on chain. Each $\kappa$ versions its evidence schema.

Fig.~\ref{fig:architecture}(b) gives the claim state machine and protocol invariants. Five sample contracts (Table~\ref{tab:kcatalog}) register against a local development chain at a mean of 552\,k gas each. The same contract suite also runs on the Base Sepolia public testnet, integrated with live Circle USDC and UMA OOv3 rather than mocks (Kleros has no Base Sepolia deployment, so we substitute a mock arbitrator). The contract suite, evaluation harness, and reproduction scripts are available at \url{https://github.com/mpi-dsg/recourse}.

\subsection[Designing a kappa]{Designing a $\kappa$}
\label{sec:designing-kappa}

To onboard a new workload, an operator follows five steps. (1)~\emph{Scope} names an (action template, resource scope) pair tight enough that the runtime can decide \textsc{WithinScope} from an EffectReceipt's resource id alone (schema prefix, repository name, parameter path prefix). (2)~\emph{Recovery operator} registers a $\rho$ that restores from a snapshot or replays an undo log. Provider-cooperation requirements (freshness, retention, callable API) become preconditions in $\kappa$. (3)~\emph{Payout schedule} chooses $P$ as a capped, closed-form function of receipt fields (per-row, per-byte, per-minute), so $P(o,r)$ computes from the contract alone. (4)~\emph{Bond size} sets bond at least as large as the expected concurrent-claim multiplier times the payout cap. RQ3 places the knee at $1\times$ cap for independent failures and the correlated-burst limit near the multiplier. (5)~\emph{Evidence schema} declares the provenance level the contract requires (L1 provider-attested, L0 replayable). \textsc{EvidenceOK} checks against that declared level.

These five steps are a one-time integration cost per (action template, scope) pair. The ongoing burden is keeping the compensator callable, the adapter's signing key custodied, and the evidence store's retention window outliving $\Delta$. A stale compensator or expired retention window degrades the receipt's provenance level (Table~\ref{tab:provenance}) rather than corrupting settlement, pushing the claim from optimistic toward arbitrated or contract-default.

\begin{table}[t]
\caption{Sample $\kappa$ catalog registered on a local development chain.}
\label{tab:kcatalog}
\centering
\scriptsize
\begin{tabular}{@{}llrrr@{}}
\toprule
\textbf{Name} & \textbf{Scope} & \textbf{$P_{\max}$} & \textbf{Bond} & \textbf{Gas} \\
\midrule
\tablekcatalogue
\bottomrule
\end{tabular}
\end{table}

\section{Evaluation}
\label{sec:evaluation}

We evaluate Recourse against five research questions, ordered by impact. RQ1 asks how much uncompensated harm Recourse removes and where it differs from centralized escrow. RQ2 checks whether the router sends each failed claim to the intended destination. RQ3 tests adversaries, bond sizing, correlated failures, and property fuzzing. RQ4 asks whether provider-compatible local sandboxes produce receipts the protocol can verify. RQ5 measures on-chain settlement cost and autonomy under different approval rates. All baselines share traces, fault injectors, and metric extractors. We run RQ1--RQ3 and RQ5 on a deterministic synthetic harness with injected faults. RQ4 is the paper's only sandbox-integration evidence, scoped to the three provider-compatible local environments described in Section~\ref{sec:rq4-provider}.

All harness, classifier, fuzzer, and contract-gas runs execute on an Apple M-series workstation (macOS, 64\,GB RAM) against a Foundry local development chain. Live-provider adapters target managed sandboxes (Postgres, GitHub, AWS Parameter Store). The harness uses seed 42, 24--60 calls per cell, and five repetitions per cell. We also cross-check headline numbers across 10 base seeds (42--51) with Student-$t$ 95\% confidence intervals over the 50 resulting sample points. The Postgres recourse contract (Fig.~\ref{fig:architecture}(c), card~\circled{1}) uses a composite payout (row-count plus downtime), a \$1{,}000 payout cap, and a bond at $10\times$ the cap. We normalize trace units as one row deleted for Postgres, one byte of unreachable git object for Git, and one parameter-version delta for Cloud. Dollar amounts in $\kappa$ are USDC. The coverage metric is the approval-free risky-call share, $N_{\text{risky,approval-free}}/N_{\text{risky}}$. Per failed action $i$, $\textsc{Recovered}_i=\max(0,G_i{-}A_i)$, $\textsc{Covered}_i=\min(A_i,P_i,B_i)$, and $\textsc{Uncompensated}_i=\max(0,A_i{-}\textsc{Covered}_i)$, with $G_i$ gross harm, $A_i$ post-recovery residual, $P_i=P(o_i,r_i)$ payout, and $B_i$ bond.

\subsection{RQ1: Loss Reduction and the Trust Model}
\label{sec:rq1}

All seven baselines see the same trace at approval rate 0.6 (Table~\ref{tab:rq1-baselines}), but their admission policies execute different subsets of it. NONE and OAP leave all 6{,}796 units uncompensated. Atomix and auth-plus-local recovery restore 1{,}796 units and leave 5{,}000. Centralized escrow and Recourse execute 5{,}464 units of gross harm, recover 1{,}796, cover 1{,}198, and leave 2{,}470.

Table~\ref{tab:rq4-ablation} isolates the cumulative ablation at approval rate 0.7: the single-seed Atomix-to-Recourse reduction is 42.0\,\% (6{,}057 to 3{,}511), which strengthens to 49.8\,\%\,$\pm$\,12.6\,pp under a 10-seed sweep (95\% CI). Recourse uses the same payout rule as centralized escrow and produces the identical loss number. Both share the payout schedule, bond cap, and fault stream, so the harness has no loss-axis separation. The distinction is on the four trust-model axes (Section~\ref{sec:settlement}): centralized escrow has externally reconstructable history, Recourse has all four flags, and OAP and Atomix have none.

\begin{table}[t]
\caption{Gross, recovered, covered, and uncompensated harm per baseline (Postgres harness, normalized synthetic units, mean of five deterministic repetitions).}
\label{tab:rq1-baselines}
\centering
\scriptsize
\begin{tabular}{@{}llrrrr@{}}
\toprule
\textbf{Workload} & \textbf{System} & \textbf{Gross} & \textbf{Rec.} & \textbf{Cov.} & \textbf{Uncomp.} \\
\midrule
\tablerqone
\bottomrule
\end{tabular}
\end{table}

\begin{table}[t]
\caption{Postgres harness ablation over normalized loss, objective-settlement share, false acceptance, and bond utilization.}
\label{tab:rq4-ablation}
\centering
\scriptsize
\begin{tabular}{@{}p{1.55cm}llll@{}}
\toprule
\textbf{Baseline} & \textbf{Uncomp.} & \textbf{Obj.\ share} & \textbf{False acc.} & \textbf{Used} \\
\midrule
\tablerqfour
\bottomrule
\end{tabular}
\end{table}

\subsection{RQ2: Routing Correctness and Rejection Reasons}
\label{sec:rq2}

We inject ten fault classes, one per predicate plus NONE, and compare the router's observed route against a pre-registered expected route (Table~\ref{tab:rq5-routing}). Within this labeled fault suite, every class routes to its expected destination at routing correctness 1.0. NONE routes to the optimistic path, and subjective payout routes to arbitration. Observability and recovery-evidence faults also route to arbitration, while boundedness faults route to exclusion. Two annotators labeled each case independently from the contract text, with initial agreement of 8/10. We resolved the two disagreements (recovery partial failure, out-of-cap loss) against the protocol rules before this run. Because we implement the router, the labels, and the predicates, this comparison measures conformance. Section~\ref{sec:rq3} covers adversarial stress.

\begin{table}[t]
\caption{Router destinations under injected predicate failures. Each row reports route share for one fault class.}
\label{tab:rq5-routing}
\centering
\scriptsize
\begin{tabular}{@{}lrrrr@{}}
\toprule
\textbf{Fault class} & \textbf{Inelig.} & \textbf{Optim.} & \textbf{Arbit.} & \textbf{Excl.} \\
\midrule
\tablerqfive
\bottomrule
\end{tabular}
\end{table}

Under nominal evidence (snapshot, affected-row count, downtime, L1 provider receipt), every admitted instance is eligible and every failure is objectively settleable. Under fault-mixed traffic (n=3{,}500), recovery-availability, evidence-schema, and loss-class each fire on 26\,\% of rejected admissions, bond-sufficiency fires on 9\,\%, and 13\,\% are admitted.

\subsection{RQ3: Robustness}
\label{sec:rq3}

\textbf{Adversaries.} Four adversaries deterministically perturb the recourse baseline (Table~\ref{tab:rq5-adv}, with NONE as the unperturbed baseline at 3{,}782 loss). In this sweep, \textsc{DISHONEST\_CLAIMANT} (inflated residual) escalates, and the dispute path slashes it (FR 1.00, shortfall 0). \textsc{ABSENT\_CHALLENGER} composes with latent inflation, increasing residual loss to 18{,}853, roughly 5$\times$ the cooperative baseline. Under \textsc{BOND\_STRESS}, the sweep adds 5.4\,\%\,$\pm$\,1.3\,pp to the baseline shortfall (10-seed CI) under a half-bond pro-rata bound. \textsc{COLLUDING\_CHALLENGERS} at 50/50 capture rate yields 0.58 false-rejection.

Collusion here captures a watcher-side dispute vote, not a flash-loan governance attack: UMA OOv3 disputes settle against a fixed per-claim USDC bond rather than a borrowable token-weighted vote, so flash-loan attacks against Augur- or reality.eth-style oracle governance do not transfer directly. ERC-792/Kleros arbitration's stake-weighted juror selection carries its own economic assumptions this sweep does not stress. These remain deterministic worst-case bounds. Full multi-party games stay open.

\begin{figure}[t]
\centering
\includegraphics[width=\columnwidth]{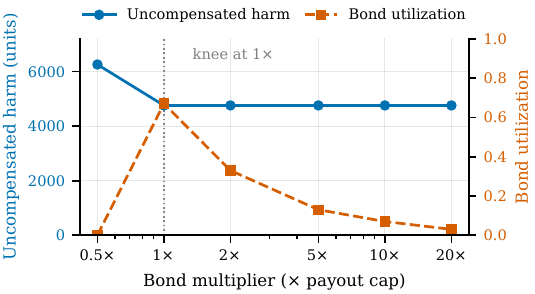}
\caption{Bond multiplier sweep: uncompensated harm plateaus at the 1$\times$-cap knee while bond utilization decays as locked capital sits increasingly idle.}
\label{tab:rq8-bondsens}
\end{figure}

\textbf{Bond multiplier.} We sweep the bond multiplier and find a threshold at 1$\times$ cap (Fig.~\ref{tab:rq8-bondsens}): an under-bonded operator (0.5$\times$ cap) leaks 6{,}260 units, while properly bonded operators plateau at 4{,}760 from 1$\times$ onward as the per-claim payout cap binds before the bond does. Bond utilization drops from 0.67 to 0.03 across the sweep.

\begin{figure}[t]
\centering
\includegraphics[width=\columnwidth]{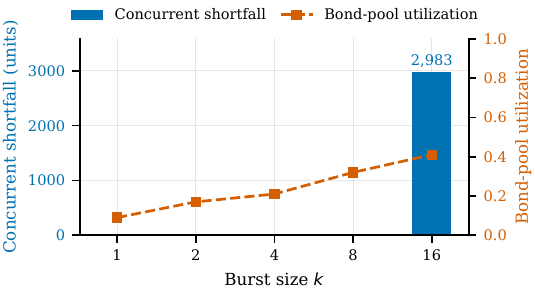}
\caption{Correlated-burst sweep ($p{=}0.3$, 10$\times$ bond): concurrent shortfall stays at zero through burst size $k{=}8$ and jumps once the shared pool saturates at $k{=}16$.}
\label{tab:rq10}
\end{figure}

\textbf{Correlated failures.} We inject burst-mode failures ($p{=}0.3$ per epoch, burst size $k$) to probe the independence assumption in the 10$\times$ bond (Fig.~\ref{tab:rq10}): the shared pool absorbs every burst with zero concurrent shortfall up to $k{=}8$, then saturates at $k{=}16$ (shortfall 2{,}983). The 10$\times$ bond therefore sizes the protocol against at most \textasciitilde 10 concurrent claims, and we recommend operators size bonds to this limit.

\textbf{Fuzzing.} We fuzz four classifier properties (idempotent classification, monotone scope, capped payout, never-crash) across 800 Hypothesis examples and find no counterexamples. Three stateless contract fuzzers pass 768 Foundry runs, and three stateful invariants (bond covers locked, monotone claim transitions, payout cap) hold across 384{,}000 calls. These three counts use different statistical units, so we report them separately rather than summed.

\begin{table}[t]
\caption{Adversary sweeps against the recourse baseline. FA/FR report false acceptance and false rejection; loss and shortfall use normalized units.}
\label{tab:rq5-adv}
\centering
\scriptsize
\begin{tabular}{@{}lrrrr@{}}
\toprule
\textbf{Adversary} & \textbf{FA} & \textbf{FR} & \textbf{Loss} & \textbf{Shortfall} \\
\midrule
\tablerqfiveadv
\bottomrule
\end{tabular}
\end{table}

\textbf{Takeaway.} Under the challenge assumption, the protocol slashes inflated claims, though \textsc{ABSENT\_CHALLENGER} and \textsc{COLLUDING\_CHALLENGERS} cause the largest loss inflation in our model. The 1$\times$ bond threshold and the $k\!\approx\!$mult correlated-burst limit are workload-specific heuristics, and general bond sizing needs a workload model.

\subsection{RQ4: Provider-Compatible Local Sandboxes}
\label{sec:rq4-provider}

The adapters and compensators drive three local, provider-compatible environments: a Docker Postgres container, a local bare Git repository, and a LocalStack SSM instance. \textbf{Postgres}: 20 cells, 15 admitted, 5 rejected on scope violations, 33 gross harm units recovering to 17 residual. \textbf{Git}: 20 cells, 14 admitted (8 reflog-recoverable, 6 gc-unrecoverable), 6 rejected, 3{,}223 gross harm units recovering to 1{,}398 residual. \textbf{Cloud parameter store}: 20 cells, 12 admitted, 8 rejected, 12 gross harm units recovering to 6 residual. The adapters sign EIP-712 receipts in every case. These environments are ephemeral, single-machine, and fully operator-controlled, so a second local process re-checking the same container is not independent verification: no party but the operator can reach the state once the run ends. The receipts are therefore \emph{L2 evidence} (Table~\ref{tab:provenance}), not L0 or L1. They establish that the adapter/compensator pipeline runs correctly against provider-compatible interfaces. Independent third-party or provider attestation is a separate claim this evidence does not make.

\textbf{Takeaway.} 19 of 60 cells (31.7\,\%) across three workloads are rejected on concrete scope violations, and admitted cells exercise both successful and failed recovery paths.

\textbf{Future work: upgrading provenance tier.} Reaching L1 requires real managed providers (a hosted Postgres service, a GitHub-hosted repository, real AWS instead of LocalStack) plus a provider-side attestation channel the adapter cannot forge alone, e.g., a cloud audit-log signature or a webhook-signed event, none of which the current adapters support. Reaching L0 requires the underlying state to be independently, publicly checkable without any cooperation from the operator. For private infrastructure like a database or a private repository, that realistically means anchoring to a public log or blockchain rather than to the provider's own API. Both are open engineering work, not a configuration change to the existing sandboxes.

\subsection{RQ5: Deployability}
\label{sec:rq5}

\textbf{On-chain end-to-end.} We measure 1.854\,M in-process call-gas (happy) and 2.060\,M (disputed) from isolated Foundry runs. Claim submission dominates at 411\,k, and contract registration is a one-time 578\,k. Table~\ref{tab:rq6-gas} lists every measured state-changing call. These are in-process EVM call-gas figures, not external transaction-receipt gas, and they exclude deployment, harness setup, token preparation, and view calls. The same Base Sepolia deployment (Section~\ref{sec:implementation}) totals 7.88\,M execution gas plus 318\,k L1 data gas across 8 transactions, against live Circle USDC and UMA OOv3.

\textbf{Live verification.} We independently verify the deployment: on-chain bytecode matches the compiled source by hash for all six contracts, confirmed without the deployer's private key, and the contracts correctly reference live Circle USDC and UMA OOv3 rather than placeholders.

\textbf{Cost in dollars.} These are illustrative conversions of the local call-gas totals above, not measured transaction costs. At ETH \$3{,}000 and typical gas prices, the per-claim settlement gas translates to sub-cent at Base Sepolia (0.011 gwei measured), \$278 happy / \$309 disputed / \$87 one-time register at Base mainnet (50 gwei), and \$167 / \$185 / \$52 at Ethereum L1 (30 gwei). These figures exclude transaction overhead, L1 data fees, UMA assertion bonds, arbitrator fees, and opportunity cost of locked collateral. The \$278 happy-path Base-mainnet cost is 28\,\% of the \$1{,}000 payout cap used in the Postgres $\kappa$ (Table~\ref{tab:kcatalog}). The \$309 disputed cost is 31\,\%. Gas cost is flat per claim while the cap scales with covered value, so overhead falls as a share of covered value for $\kappa$ with higher cap.

\textbf{Capital.} The bond sizes to the per-claim payout cap: locked bond averages 36 units at 0.67 utilization on Postgres, while residual loss outside the payout schedule appears as shortfall (4{,}595). The bond guarantees the contracted payout. Losses above the cap remain outside the guarantee, and under-bonding beyond 1$\times$ cap directly leaks loss (RQ3). Postgres meets the 25\,ms per-call latency budget against a 250\,ms ceiling, with sandbox-measured latency at 6.46\,ms.

\textbf{Approval-rate sweep.} Across approval rates 0--1, recourse and centralized share the lowest loss curve (loss-identical, see RQ1) while OAP and the uncovered baseline run roughly 1{,}200 units higher, and human approval traces the expected coverage--loss trade-off.

\begin{table}[t]
\caption{Median Foundry in-process call gas per operation. Settle/rule rows include the optimistic-oracle or arbitrator callback that finalizes the claim and pays out.}
\label{tab:rq6-gas}
\centering
\scriptsize
\begin{tabular}{@{}lrr@{}}
\toprule
\textbf{Tx} & \textbf{Happy} & \textbf{Disputed} \\
\midrule
\tablerqsix
\bottomrule
\end{tabular}
\end{table}

\subsection{Threats to Validity}
\label{sec:threats}

The synthetic harness in RQ1--RQ3 and RQ5 exercises the real classifier, baseline contracts, and bond accounting in deterministic normalized units. RQ4 and the on-chain end-to-end cover only the local-sandbox and local-chain tiers, not managed infrastructure. Bond economics assume a $10\times$ payout-cap bond that covers independent failures up to ten cap-sized claims. Correlated workload-wide failures need claim priority, pro-rata payout, or external surety. Runtime trust remains centralized in one operator that runs the classifier, adapter, compensator, and evidence store, and we leave multi-party designs to future work. The feature-matched controls approximate autonomy-only, local-compensation-only, and centralized-escrow deployments. The centralized-escrow baseline matches the payout rule and dispute window of commercial marketplace-escrow products, while deployment-level differences in KYC, settlement cadence, and chargeback exposure remain out of scope.

The Base Sepolia deployment (Section~\ref{sec:rq5}) validates correctness and gas cost, not mainnet adversarial pressure, real collateral economics, or contested-dispute volume, and the three provider adapters validate against local sandboxes, not production traffic. Watcher participation under real economic stakes, adapter operation against live provider accounts, and dispute behavior under genuinely contested claims remain open before a production deployment.

\section{Related Work}
\label{sec:related}

Recourse settles residual harm from a concrete admitted tool action after attempted recovery, distinct from \emph{algorithmic recourse} in interpretable ML~\cite{Wachter2018CounterfactualExplanations,Ustun2019ActionableRecourse,Karimi2021AlgorithmicRecourse}.

\textbf{Pre-action and compensation.} OAP~\cite{Uchibeke2026BeforeToolCall} gates a tool call before execution. Atomix~\cite{Mohammadi2026Atomix} and Sagas~\cite{GarciaMolina1987Sagas} resolve workflows on abort with local rollback. ACRFence~\cite{Zheng2026ACRFence} prevents replay hazards on checkpoint restore. Residual harm after an admitted action stays out of all four.

\textbf{Payment, insurance, standards.} TessPay~\cite{Goenka2026TessPay}, ARS~\cite{Hua2026QuantifyingTrust}, Insured Agents~\cite{Hu2025InsuredAgents}, AP2~\cite{Parikh2025AgentPaymentsProtocol}, ERC-8004~\cite{DeRossi2025ERC8004}, and MCP~\cite{ModelContextProtocol2025Authorization} settle task payment, transactions, trust, identity, or authorization. Agent Contracts~\cite{Ye2026AgentContracts} covers a wider object. SLA-compensation smart contracts~\cite{Scheid2019DynamicSLACompensation} pay out on service-level breach. Recourse settles action-level harm, not SLA KPIs.

\textbf{Bonded protocols.} The optimistic-bond-with-challenge pattern originates in Truebit~\cite{Teutsch2023ScalableVerification} and scales in rollup fraud proofs~\cite{Kalodner2018Arbitrum,OPLabsFaultProof}. Recourse consumes UMA OOv3~\cite{UMA2026Oracle,UMA2026CustomBondLiveness}, Augur~\cite{Peterson2020Augur}, reality.eth~\cite{RealityEth2017Whitepaper}, and Kleros~\cite{Kleros2026Escrow,Kleros2026Oracle} via ERC-792/1497~\cite{Lesaege2017ERC792,Vitello2018ERC1497}. Nexus Mutual~\cite{Karp2018NexusMutual} settles protocol-failure, not action-level, harm. Verifiable TLS-attestation oracles (DECO~\cite{Zhang2020DECO}, Town Crier~\cite{Zhang2016TownCrier}) can synthesize L1 receipts when providers do not natively sign. Receipts inherit tamper-evidence guarantees from classical audit-log and certificate-transparency design~\cite{SchneierKelsey1999AuditLogs,Laurie2013RFC6962}.

\section{Conclusion}
\label{sec:conclusion}

Recourse targets one narrow failure mode: bounded residual loss after an admitted autonomous action, binding it to scope, recovery, evidence, payout, and collateral. It cuts uncompensated harm substantially in the Postgres harness, routes injected predicate failures to their expected destinations, and settles on-chain within a bounded gas budget. The Base Sepolia deployment demonstrates contract registration, bonding, and failure routing via UMA/Kleros. Multi-party games, broader workloads, and insurer-backed surety remain open.

\section*{Acknowledgment}
The authors used AI tools for editorial assistance: drafting, copy-editing, and trimming for length. All technical content, results, and conclusions are the authors' own.

\IEEEtriggeratref{19}
\bibliographystyle{IEEEtran}
\bibliography{references}

\end{document}